\documentclass[]{aastex701}
\usepackage{graphicx}
\usepackage{placeins}  
\usepackage{hyperref}
\usepackage[utf8]{inputenc}
\usepackage{newunicodechar}
\newunicodechar{ }{~}  
\usepackage{epstopdf}

\begin{document}

\title{Observational Evidence for Counter-helicity Magnetic Reconnection in a Solar Eruption}

\author[orcid=0009-0006-6772-075X,gname=China]{Jinrui Chang}
\affiliation{Yunnan Observatories, Chinese Academy of Sciences, Kunming 650216, People's Republic of China}
\affiliation{University of Chinese Academy of Sciences, Beijing, 100049, People's Republic of China}
\email{changjinrui@ynao.ac.cn}  

\author[orcid=0000-0002-5302-3404,gname=China]{Yi Bi}
\affiliation{Yunnan Observatories, Chinese Academy of Sciences, Kunming 650216, People's Republic of China}
\affiliation{Yunnan Key Laboratory of Solar Physics and Space Science, 396 Yangfangwang, Guandu District, Kunming 650216, PR China}
\email[show]{biyi@ynao.ac.cn}

\author[orcid=0000-0002-1322-9061,gname=China]{Bo Yang}
\affiliation{Yunnan Observatories, Chinese Academy of Sciences, Kunming 650216, People's Republic of China}
\email{yangbo@ynao.ac.cn}

\author[orcid=0000-0002-3804-7395,gname=China]{Junchao Hong}
\affiliation{Yunnan Observatories, Chinese Academy of Sciences, Kunming 650216, People's Republic of China}
\email{hongjunchao@ynao.ac.cn}

\author[orcid=0000-0003-3462-4340,gname=China]{Jiayan Yang}
\affiliation{Yunnan Observatories, Chinese Academy of Sciences, Kunming 650216, People's Republic of China}
\email{yjy@ynao.ac.cn}

\author{Qingmei Wang}
\affiliation{Yunnan Observatories, Chinese Academy of Sciences, Kunming 650216, People's Republic of China}
\email{1936825096@qq.com}

\begin{abstract}

Magnetic reconnection between coronal magnetic systems carrying opposite self-helicity may play a role in solar eruptions, but  observational evidence remains limited. We investigate an M7.0 flare in NOAA Active Region 13615 on 2024 March 28 using multiwavelength observations and nonlinear force-free field extrapolations. The reconstructed coronal field reveals a low-lying positive-helicity core field beneath an overlying magnetic system of opposite sign. During the eruption, the footpoint connectivity of these two magnetic systems changes markedly: field lines rooted in the western footpoint region change from positive to negative helicity, and the positive-helicity domain is substantially reduced. These changes are accompanied by a remote chromospheric brightening, intermittent EUV stripe-like brightenings extending from the source region toward the remote chromospheric brightening, the subsequent formation of large-scale coronal loops, and a weak outer hard X-ray source located at a footpoint of the core field. Together, these results suggest that the eruption was closely associated with reconnection between the core field and the overlying counter-helicity system, providing observational evidence that counter-helicity reconnection can contribute to the destabilization of eruptive solar magnetic fields.

\end{abstract}



\section{Introduction} 

Solar flares are the most energetic events in the solar system, releasing $10^{28}$ to $10^{32}$~erg through magnetic reconnection in active regions (ARs), often accompanied by coronal mass ejections (CMEs) that impact space weather. Strong flares are statistically linked to complex ARs with helical magnetic configurations \citep{Toriumi2019, Liu2023}. Magnetic helicity, quantifying the twist, shear, and linkage of field lines, measures this complexity and is conserved even during resistive processes like flares \citep{Berger1984}. Continuous helicity injection with a predominant sign can accumulate free energy, exceeding stability thresholds and triggering eruptions \citep{Moon2002, Zhang2005, Romano2011, Tziotziou2012, Jasinski2025, Sun2025}. These events frequently involve the destabilization and eruption of magnetic flux ropes (MFRs), twisted magnetic structures that store free magnetic energy \citep{Priest2002, Shibata2011, Jiang2023}. Hot channels, observed in extreme ultraviolet (EUV) wavelengths by instruments such as the Atmospheric Imaging Assembly (AIA) on the Solar Dynamics Observatory (SDO), serve as observational proxies for MFRs \citep{Zhang2012, Cheng2012, Song2015}. These elongated, high-temperature plasma structures signal the presence of an underlying MFR that can erupt to form a CME, providing insights into eruption initiation and evolution.

Observational evidence for counter-helicity reconnection in major solar eruptions remains limited, although a few relevant examples have been reported. In the X10 flare on 29 October 2003, vector magnetograms revealed a complex AR with mixed helicity signs: the main positive-polarity sunspots were dominated by negative helicity but contained interspersed patches of positive helicity \citep{Liu2007}. After the flare, the average current helicity density decreased by $\sim$50\% in the main sunspots, with positive helicity patches decaying proportionally or more. Hard X-ray footpoints coincided with these disappearing positive patches, suggesting that the flare was triggered by reconnection between counter-helical flux systems, with the helicity reduction attributed to ejection via the associated halo CME \citep{Liu2007}. However, not all counter-helicity scenarios lead to eruptions. Active regions with successive injection of opposite-sign helicity exhibit sign changes in helicity flux, which prevent the accumulation of excess helicity and the formation of highly twisted magnetic flux ropes, resulting in sub-C-class flares without CMEs \citep{Vemareddy2021, Vemareddy2022}.  Similarly, \citet{Hou2023} reported a pre-eruption double-decker filament composed of two vertically stacked MFRs with opposite magnetic twist, in which only the upper branch erupted while the lower branch remained. These cases indicate that the presence or injection of opposite-sign helicity alone is insufficient for major energy release unless a favorable coronal magnetic connectivity and geometry permit direct interaction between the two magnetic systems. Identifying such eruptive configurations therefore requires detailed knowledge of the coronal magnetic field, obtained through magnetic field extrapolation or direct observational diagnostics.

A related product of reconnection between twisted flux systems is the counter-helical magnetic flux rope (CHFR) identified by \citet{Jia2024a}. Their simulations suggest that reconnection between interlaced MFRs carrying opposite helicity tends to generate a new flux rope containing segments with opposite helicity signs along its axis. Because the CHFR is non-force-free and unstable, its subsequent relaxation may lead to gradual annihilation of the opposite-helicity components and associated magnetic-energy release. Direct observational identification of such structures remains uncertain. Multi-spacecraft measurements by \citet{Rodríguez-García2022} inferred opposite helicity from different portions of the same ICME, providing a possible interplanetary indication of a CHFR. In the solar atmosphere, \citet{Jacob2025} reported coexisting oppositely signed twist in a non-eruptive prominence–cavity system. This structure may represent a solar analogue of a counter-helical magnetic configuration.

MHD simulations by \citet{Linton2001} show that magnetic reconnection between two twisted flux ropes can undergo merging, slingshot, bounce, or tunnel reconnection, depending on the sign of the twist (same or opposite), the twist magnitude, and the contact angle between interacting MFRs. Interactions between tubes with a contact angle greater than $90^\circ$ can lead to slingshot reconnection, which radically alters tube connectivity and efficiently releases energy. This can occur between two counter-helicity tubes with high twist and has also been reported between two same-helicity tubes with low twist \citep{Linton2006, Jiang2013}. Field structures with the same helicity sign may undergo merging reconnection, producing a merged structure with relatively limited energy release. Merging reconnection occurs between same-helicity, small-angle filament threads \citep{Bi2012}.

More generally, the interaction of adjacent current channels or flux structures depends on the relative alignment of their currents. In many cases, co-aligned currents are associated with the coalescence instability, which drives mutual attraction and merging, whereas counter-aligned currents favor the tilt instability, which leads to rotation and separation \citep{Keppens2014, Liu2020, Knoll2006, Pritchett2007}. Although the global evolution of counter-aligned current channels is repulsive, their deformation can build strong current layers at the interaction interface, where localized magnetic reconnection may still occur. This provides a useful physical context for interpreting small-angle interactions in counter-helicity systems.

Tube reconnection refers to a specific case of magnetic reconnection involving the reconfiguration of two entire interacting flux tubes. When two flux tubes approach each other at a small contact angle, their interaction may nevertheless manifest locally as small-angle magnetic reconnection between the field lines that compose them. A related situation occurs in braided coronal fields, where reconnection develops between marginally misaligned magnetic strands, as envisioned by  \citet{Parker1988}. In this model, the stressing and tangling of magnetic field lines by photospheric motions ultimately lead to many small-angle reconnection events, releasing energy and producing transient outflows. Observations of nanojets---short-lived, small-scale jets oriented perpendicular to the axis of coronal loops---provide direct evidence of such braiding-induced reconnection occurring in apparently quiescent loops, as reported by \citet{Antolin2021}. These nanojets are interpreted as the observational signature of localized, small-angle reconnection between marginally misaligned magnetic strands within a braided field. Similar small-angle reconnection processes have been widely identified, for example, through the detections of nanojet-like ejections perpendicular to solar tornadoes \citep{Chen2017}, flaring loops \citep{Wang2025}, double-decker filaments \citep{Liu2025}, and tangled superpenumbral fibrils \citep{Chen2025}.

In this paper, we investigate an eruption associated with an M7.0 flare. By combining multiwavelength observations with nonlinear force-free field extrapolations, we show that the eruption was closely associated with magnetic reconnection between the MFR and the overlying magnetic field with opposite magnetic helicity.



\section{Data} \label{sec:style}

In this study, we primarily utilize data from the AIA and the New Vacuum Solar Telescope \citep[NVST;][]{Liu2014}. Additionally, we incorporate data from the HMI on SDO, full-disk spectral scans in H$\alpha$ from the Chinese H$\alpha$ Solar Explorer (CHASE), soft X-ray flux from the Geostationary Operational Environmental Satellite (GOES), and hard X-ray data from the Hard X-ray Imager (HXI) on the Advanced Space-based Solar Observatory (ASO-S).

The AIA aboard the SDO \citep{Pesnell2012} obtains full-disk images in ten ultraviolet and extreme-ultraviolet passbands with a spatial sampling of $0.6''$ per pixel and a cadence of 12 seconds. AIA extreme ultraviolet channels sample plasmas at characteristic temperatures: 131~\AA\ ($\sim$10~MK and 0.6~MK), 94~\AA\ ($\sim$7.2~MK), 335~\AA\ ($\sim$2.5~MK), 211~\AA\ ($\sim$1.9~MK), 193~\AA\ ($\sim$1.5~MK), 171~\AA\ ($\sim$0.9~MK), and 304~\AA\ ($\sim$0.05~MK) \citep{Lemen2012}.

The NVST provides H$\alpha$ images with a central wavelength of 6562.8~\AA\ and a bandwidth of 0.25~\AA. These images have a field of view of $185'' \times 185''$, a spatial sampling of $0.168''$ per pixel, and a temporal resolution of 41.3 seconds. We processed the data primarily for alignment and co-registration following \citet{Ji2019}. Specifically, the NVST H$\alpha$ images are aligned with the AIA 304~\AA\ channel images, as both are sensitive to chromospheric activity.

Additional data products were used to supplement the analysis. The HMI on SDO \citep{Schou2012} provides line-of-sight magnetograms (45-second cadence, $0.5''$ per pixel) and vector magnetic field data (12-minute cadence). The CHASE/HIS (H$\alpha$ Imaging Spectrograph; \citealt{Liu2022}) obtains full-disk spectral scans in H$\alpha$ (6559.7--6565.9~\AA) and Fe~{\sc I} (6567.8--6570.6~\AA) bands; the processed data \citep{Qiu2022} have a spatial sampling of $1.05''$ per pixel and a cadence of $\sim$71 seconds. GOES data provide soft X-ray flux in the 0.5--4.0~\AA\ and 1--8~\AA\ bands for flare classification, among other space weather parameters. The HXI on ASO-S images solar flares in the 30--200~keV energy range with high spatial, temporal, and spectral resolution \citep{Gan2019, Su2019}.

\section{Results} \label{sec:floats}
\subsection{Analysis of the eruption} \label{subsec:tables}
Active Region (AR) NOAA 13615, located at solar coordinates S14W51, was observed jointly by the SDO, NVST, and CHASE. On March 28, 2024, an eruption occurred in this AR, which had a complex $\beta\gamma\delta$ magnetic configuration. The magnetic field within the AR exhibited a highly mixed polarity structure, characterized by a large positive polarity region in the northeast and an interwoven, ribbon-like distribution of mixed polarities in the center. According to GOES soft X-ray observations, the associated flare was classified as an M7.0 event, with its onset at 06:16 UT and peak at 06:29 UT.

Figure 1 presents multiwavelength observations depicting the temporal evolution of the filament eruption, including AIA observations in the 94~\AA, 131~\AA, 211~\AA, and 304~\AA\ channels. NVST H$\alpha$ $-0.6$~\AA\ images (Figures 1a--1b) show that the filament structures within the red box weakened between 06:18 UT and 06:22 UT. At 06:22 UT, double flare ribbons emerged in the NVST H$\alpha$ line center, signifying the onset of pinch-off reconnection in the overlying coronal magnetic field. Concurrently, a distinct chromospheric brightening appeared south of these ribbons. This unusual brightening expanded rapidly southwestward at a projected speed of up to 167~$\rm km~s^{-1}$, as shown by the time–distance plots in Figures 2a and 2b constructed along the blue slit in Figure 1d. By 06:26 UT, it covered an area of approximately 15~$\rm Mm^2$ . The brightening was also clearly observed in coronal images from SDO/AIA.

At 06:24 UT, the formation of a hot channel was observed in AIA 94~\AA\ images. Figures 1g and 1h illustrate the morphology and motion of this hot channel, which propagated southwestward at approximately 55~$\rm km~s^{-1}$, as indicated by the AIA 94 \AA\ time-distance plot (Figure 2c). In AIA 211 \AA\ images, we also identified prominent stripe-like structures extending between the eruption source region and the remote chromospheric brightening, as shown in Figures 1e, 1j, and 1o. Unlike well-defined coronal loops, these structures consist of intermittent compact brightenings and short bright segments rather than continuous loop-like emission. Their intermittent occurrence is more clearly seen in the accompanying animation of Figure 1, where similar structures are also visible in the AIA 304~\AA\ and NVST H$\alpha$ passbands. These stripe-like structures appeared during 06:23:43–06:26:19 UT, when the remote brightening observed in both EUV and H$\alpha$ wavelengths rapidly expanded (Figures 2a and 2b).  Their close spatial and temporal association suggests that the stripe-like structures are related to the remote brightening. The remote brightening may be driven by ejections originating from the eruption source region, with the observed stripe-like structures tracing visible portions of these ejections.

Subsequently, after 06:40 UT, a large-scale coronal loop system, rooted in the chromospheric brightening region, became visible in Figures 1j and 1n and was likely formed through chromospheric evaporation \citep{Acton1982, Huang2020}. This loop system crossed the PIL, marked by the yellow line in Figures 1k and 1n, along which the filament was located, with a spatial scale far exceeding that of the post-flare loop system associated with the double ribbons.

\subsection{Results from NLFFF extrapolation} \label{subsec:tables}

To investigate the formation mechanism of this event, we reconstructed the coronal magnetic field by applying the nonlinear force-free field (NLFFF) extrapolation method developed by \citet{Jarolim2023} to vector magnetograms from the HMI. Additionally, we calculated a field-line self-helicity proxy following \citet{Berger2006}. The total helicity proxy of a field line is defined as:
\begin{equation}
H = h \cdot \Phi^{2} = (T + W) \cdot \Phi^{2}
\end{equation}
where
\begin{equation}
W = \frac{1}{4\pi} \oint \oint \hat{T}(s) \times \hat{T}(s') \cdot \frac{\mathbf{x}(s) - \mathbf{x}(s')}{|\mathbf{x}(s) - \mathbf{x}(s')|^3} \, ds \, ds'
\end{equation}
is the writhe of a magnetic field line, and
\begin{equation}
T = \frac{1}{2\pi} \oint \hat{T}(s) \cdot \hat{V}(s) \times \frac{d\hat{V}(s)}{ds} \, ds
\end{equation}
is its twist. Both W and T are dimensionless quantities.

Although Equation (1) includes both the twist and writhe contributions of an individual field line, it should not be regarded as equivalent to the total relative magnetic helicity of the magnetic field \citep{Yeates2018}. In particular, compared with the relative magnetic helicity framework, the quantity derived from Equation (1) does not explicitly include the mutual helicity between different magnetic flux systems. Previous studies have therefore suggested that such a quantity is more appropriate for characterizing the self-helicity of flux-rope-like structures than for describing the full helicity budget of the coronal field \citep{Guo2017}.

In the present work, this distinction is acceptable because our main purpose is not to derive the total relative magnetic helicity, but to identify magnetic structures with different self-helicity signs and to investigate their possible interaction during the eruption. In this sense, the analysis based on Equation (1) is mainly concerned with the interaction between magnetic flux systems of different self-helicity. A more complete treatment that explicitly includes mutual helicity is beyond the scope of this study and deserves future investigation.

Figures 3c–f show the photospheric distributions of the radial magnetic field, writhe, twist, and helicity proxy, respectively, within the region of interest marked by the blue box in Figure 3a. A comparison of Figures 3d and 3e shows that the twist component is much stronger than the writhe component, although the two exhibit broadly similar spatial patterns. Consequently, the helicity-proxy distribution in Figure 3f is dominated by twist, with positive and negative patches aligned along the PIL.

We present the helicity distributions in two vertical cross-sections shown in Figures 4a and 4b, which correspond to the red and blue lines in Figure 3b, respectively. The cross-section marked by the red line is oriented approximately perpendicular to the PIL, whereas that marked by the blue line is approximately parallel to the PIL. 

In both cross-sections, a positive-helicity domain, corresponding to the eruptive core field, is clearly present. In Figure 4a, negative-helicity domains are found above and to the south of this positive-helicity domain, whereas in Figure 4b the region directly above it is almost entirely occupied by negative helicity. The helicity distributions at 06:24 UT and earlier are broadly similar in both panels. By contrast, at 06:36 UT and afterward, the helicity pattern in the cross-sections changes abruptly. A comparison of the pre-eruption and post-eruption cross-sections further shows that the positive-helicity domain contracts markedly after the eruption, while the surrounding negative-helicity domain correspondingly expands. This evolution suggests that the eruption was accompanied by a rapid restructuring of the counter-helicity magnetic systems.

To relate the cross-sectional helicity evolution to the magnetic connectivity, field lines were traced from the vertical cross-section along the red line in Figure 3b. Figure 5 shows the photospheric projection of these field lines, allowing the evolution of their footpoints and helicity signs to be examined directly. Prior to the eruption, some field lines rooted in the western footpoint region exhibit positive helicity and extend mainly along the PIL. After the eruption, at 06:48 UT, field lines rooted in the same region exhibit negative helicity and markedly different connectivity. Correspondingly, negative-helicity footpoints appear within a region previously occupied by the positive-helicity core field. Although the detailed morphology of the extrapolated field lines varies among the five analyzed times, their eastern footpoints remain rooted predominantly in the same bottom-boundary region. The persistence of this eastern footpoint region across the five analyzed times provides qualitative support for its inferred location, although the precise field-line paths remain uncertain because of limitations in the NLFFF reconstruction and field-line tracing. Taken together, these NLFFF results favor an interpretation in which the eruption involves magnetic reconnection between two closed magnetic systems of opposite helicity, namely the underlying positive-helicity core field and the overlying negative-helicity arcade.

\subsection{Hard X-ray Source Analysis} \label{subsec:tables}

Figures 1i and 3b show three hard X-ray sources near the neutral line, labeled Source 1, Source 2, and Source 3 from east to west, respectively. Source~2 is the dominant footpoint source of the main flare loop, although the two conjugate footpoints are likely unresolved because of their close spacing and the limited imaging resolution. Source~3 is located above the bright arcade and may be interpreted as a loop-top source; however, its proximity to the remote brightening region outlined in Figure 1i suggests that it may instead be related to energy release in the remote region. Source 1 is spatially associated with the relatively stable eastern footpoint region identified in the NLFFF extrapolations, where field lines belonging to the two opposite-helicity systems are rooted (Figure 5b).

As illustrated by the hard X-ray light curves (Figure 6), Source 1 is weaker than Sources 2 and 3 and exhibits two distinct enhancements. The first occurs near 06:22 UT, followed by a clear decline around 06:26–06:28 UT, and then a second, weaker enhancement appears afterward. Thus, unlike the main hard X-ray sources, Source 1 does not show sustained emission.

\section{Summary and discussion}
We present a detailed analysis of the M7.0 flare SOL2024-03-28T06:29 in NOAA AR 13615. The event was accompanied by the eruption of a hot-channel structure, followed by the appearance of a remote chromospheric brightening and the subsequent formation of large-scale coronal loops. NLFFF extrapolation reveals two distinct magnetic systems of opposite helicity along the PIL. During the eruption, the NLFFF results show that field lines belonging to these two opposite-helicity systems undergo a marked change in connectivity. In particular, field lines rooted in the western footpoint region are predominantly associated with positive helicity before the eruption and with negative helicity afterward, and the positive-helicity domain is substantially reduced. Our preferred interpretation is that the event is associated with magnetic reconnection between two closed magnetic systems of opposite helicity, namely the underlying positive-helicity core field and the overlying negative-helicity arcade. A schematic illustration of the inferred magnetic configuration and reconnection scenario is shown in Figure 7.

One potential explanation for the remote chromospheric brightening is interchange reconnection between the erupting flux rope and large-scale ambient fields, which has been proposed to produce remote brightenings and filament material leakage \citep{Yan2020}. However, this scenario is inconsistent with our observations. First, the remote brightening is located far from the eruption source region. In interchange reconnection, newly formed loops typically have one footpoint near the source and the other at the remote site \citep{Wang2023}. Second, chromospheric brightenings from such reconnection are often sporadic and confined to strong-field regions, whereas our observed brightening forms a coherent, extended patch that is not confined to regions of strong magnetic field. 

The combined observations favor a scenario in which the remote chromospheric brightening is associated with magnetic reconnection between the underlying positive-helicity core field and the overlying negative-helicity arcade. The NLFFF extrapolation reveals two closed magnetic systems approaching each other at a small projected angle in the eruption source region. Such small-angle reconnection between the two magnetic systems may produce ejections that propagate approximately transverse to the local magnetic field, thereby transporting energy toward the remote region and contributing to the chromospheric brightening. The observed stripe-like structures provide supporting evidence for this possibility. Although only a limited number of such structures are detected, they likely correspond to locations where energetic particles or heated plasma encounter relatively dense plasma and deposit sufficient energy to produce observable emission, whereas most of the propagation path remains invisible because the plasma density is too low for efficient radiation. In this picture, the majority of energetic particles or heated plasma continue to propagate and ultimately deposit their energy in the dense chromosphere of the remote region, leading to the observed chromospheric heating \citep{Polito2018}.

The local onset of magnetic reconnection is determined primarily by the magnetic shear and antiparallel field components at the interaction interface, rather than directly by magnetic helicity. Nevertheless, the opposite helicity signs indicate that the core field and overlying arcade belong to two distinct magnetic systems and may affect the large-scale outcome of their interaction. As shown in simulations by \citet{Linton2001}, field structures with the same helicity sign tend to undergo merging reconnection, in which post-reconnection magnetic tension pulls the tubes together. In contrast, counter-helicity tubes can lead to slingshot reconnection, which radically alters tube connectivity. For small crossing angles, this tension naturally drives outflows perpendicular to the tubes. Notably, \citet{Linton2001} found that counter-helicity tubes at angles $\le 45^\circ$ result in weak energy release via bounce reconnection, while slingshot reconnection occurs at larger angles, such as $\ 90^\circ$. However, their simulations used highly twisted tubes (twist parameter q = 10), which suppress reconnection at small angles. Subsequent work indicates that slingshot reconnection is favored in low-twist tubes \citep{Linton2005}, consistent with our extrapolated fields showing low or moderate twist.

Source 1, detected in hard X-rays by the HXI on board ASO-S, provides complementary evidence for energetic-particle transport to the eastern footpoint region of the interacting magnetic systems. This HXI source is spatially separated from the main flare-loop sources and is associated with the area in which field lines of opposite helicity are rooted. Its emission is weaker than that of Sources 2 and 3 and exhibits two distinct peaks. Thus, Source 1 is not continuously sustained in the same manner as the main hard X-ray sources. Its remote location and relatively weak emission resemble the outer hard X-ray sources detected at the anchor points of erupting filaments or MFRs \citep{Chen2020, Stiefel2023}. In the standard explanation, flare-accelerated electrons propagate upward into the erupting structure and subsequently precipitate at its anchor points \citep{Shibata1995}, which may account for part of the Source 1 emission. This possibility remains speculative. The later enhancement may indicate a later episode of energy release associated with helicity annihilation initiated by disturbances propagating along the reconnected magnetic arch, as modeled for counter-helical flux-rope systems by \citet{Jia2024b}. 

The temporal evolution further suggests that reconnection between the two counter-helicity systems contributed to the destabilization of the flux rope indicated by the eruptive hot channel. The remote brightening developed before the rapid acceleration of the hot channel, indicating that magnetic restructuring had already begun. The NLFFF results also suggest that the magnetic flux of the overlying arcade was smaller than that of the underlying core field. Their interaction may therefore resemble reconnection between unequal flux systems, in which a substantial fraction of the weaker overlying flux reconnects while most of the stronger core field remains intact \citep{Linton2006}. Removal or restructuring of the overlying flux would reduce the downward magnetic tension and could bring the core field closer to the threshold for torus instability \citep{Kliem2006,Hassanin2022}. In this functional sense, the process resembles the removal of overlying confinement in the breakout model \citep{Antiochos1999,Shen2012}, although the topology inferred here consists of two closed magnetic systems. It differs from tether-cutting reconnection beneath the MFR \citep{Moore2001, Jiang2021} and from eruption triggering through rapid magnetic-flux emergence \citep{Chen2000}. The observations therefore support counter-helicity reconnection as a plausible contributor to the weakening of the overlying confinement and the subsequent MFR destabilization, rather than uniquely establishing it as the eruption trigger.

Our multiwavelength observations and NLFFF modeling suggest that the solar eruption was closely associated with magnetic reconnection between the underlying flux rope and the overlying arcade of opposite helicity. Reconnection between magnetic systems carrying opposite helicity merits further investigation through MHD simulations, particularly for low-twist flux tubes in which small-angle interactions may occur efficiently. More reliable validation of this scenario will require coronal magnetic-field extrapolations that are better constrained by chromospheric vector magnetic fields \citep{Chifu2017, Fleishman2019, Kawabata2020, Schad2024, Wiegelmann2021, Metcalf1995} and, where available, direct coronal magnetic-field measurements \citep{Yang2020}.

\FloatBarrier

\begin{figure*}
\plotone{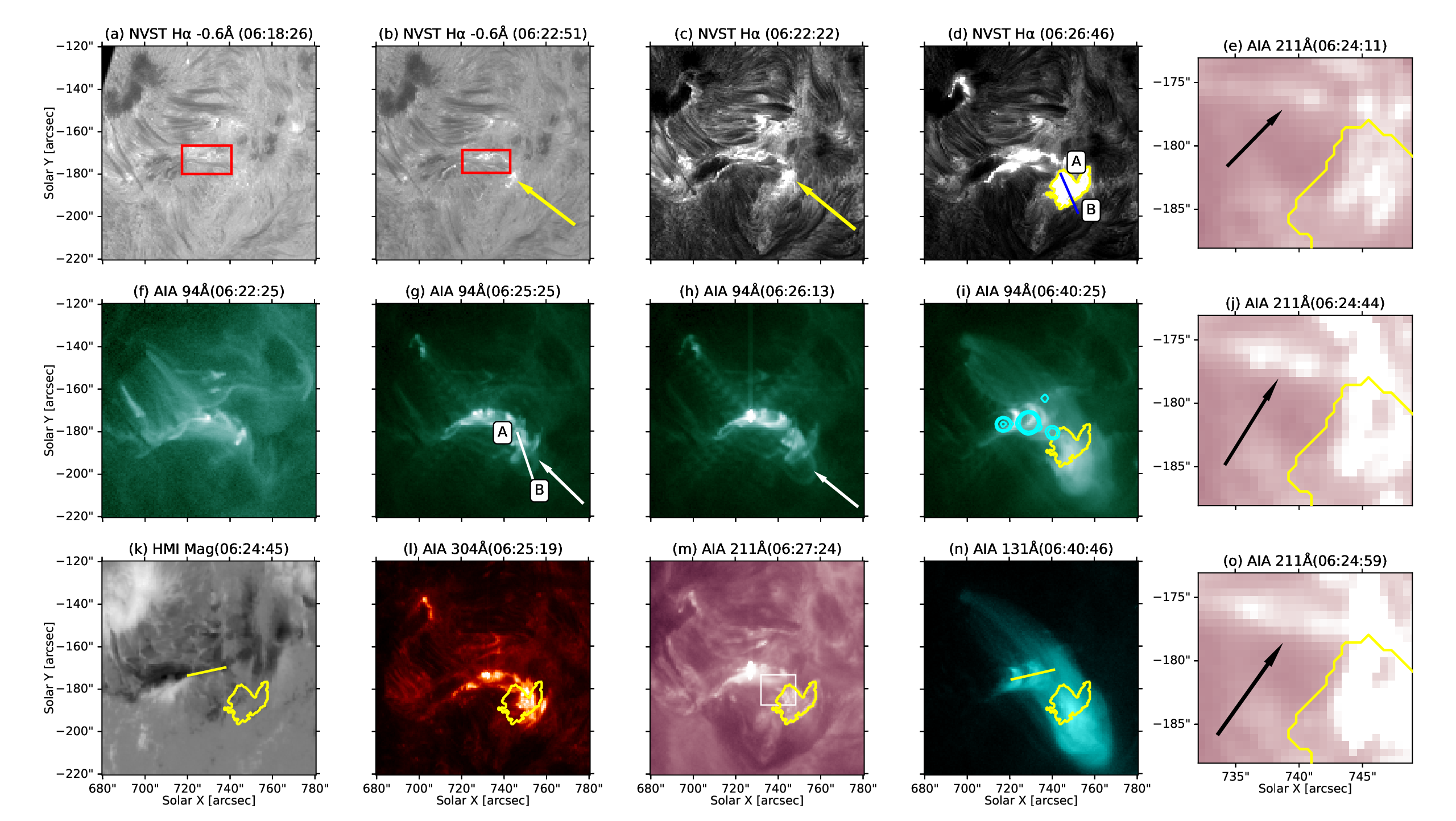}
\caption{Multiwavelength evolution of the eruption. (a)–(b) NVST H$\alpha$ blue-wing images at $-0.6$\AA; the red box encloses the filament features. (c)–(d) NVST H$\alpha$ line-center images. The yellow contour in panel (d) outlines the brightening region observed in H$\alpha$ at 06:26 UT; the same contour is overplotted in panels (e) and (i)–(o). (e), (j), (m) and (o) show SDO/AIA 211~\AA\ images. (f)–(i) SDO/AIA 94~\AA\ images. In panel (i), HXI contours in the 20--30~keV band, integrated from 06:28~UT to 06:30~UT, are overlaid and plotted from 98\% to 100\% of the peak intensity, with three contour levels separated by 1\%. (k) SDO/HMI line-of-sight magnetogram. (l) SDO/AIA 304~\AA\ images. (n) SDO/AIA 131~\AA\ image. The PIL is marked by the yellow line in panels (k) and (n).
Panels (e), (j), and (o) are enlarged views of the white box in panel (m); all other panels share the same field of view. The yellow arrow marks the initial remote brightening, the white arrows indicate the eruptive hot channel, and the black arrows mark the stripe-like brightenings. The white line AB in panel (g) marks the slit used for the time–distance plots in Figure 2(a)–(c). An accompanying animation of panels (a), (c), (e), (f), (j), (k), and (l) is available as supporting information. It covers 25 minutes of observations beginning at 06:22:07 UT on 2024 March 28 and has a duration of 4~s.}
\label{fig:1}  
\end{figure*}

\begin{figure*}
\plotone{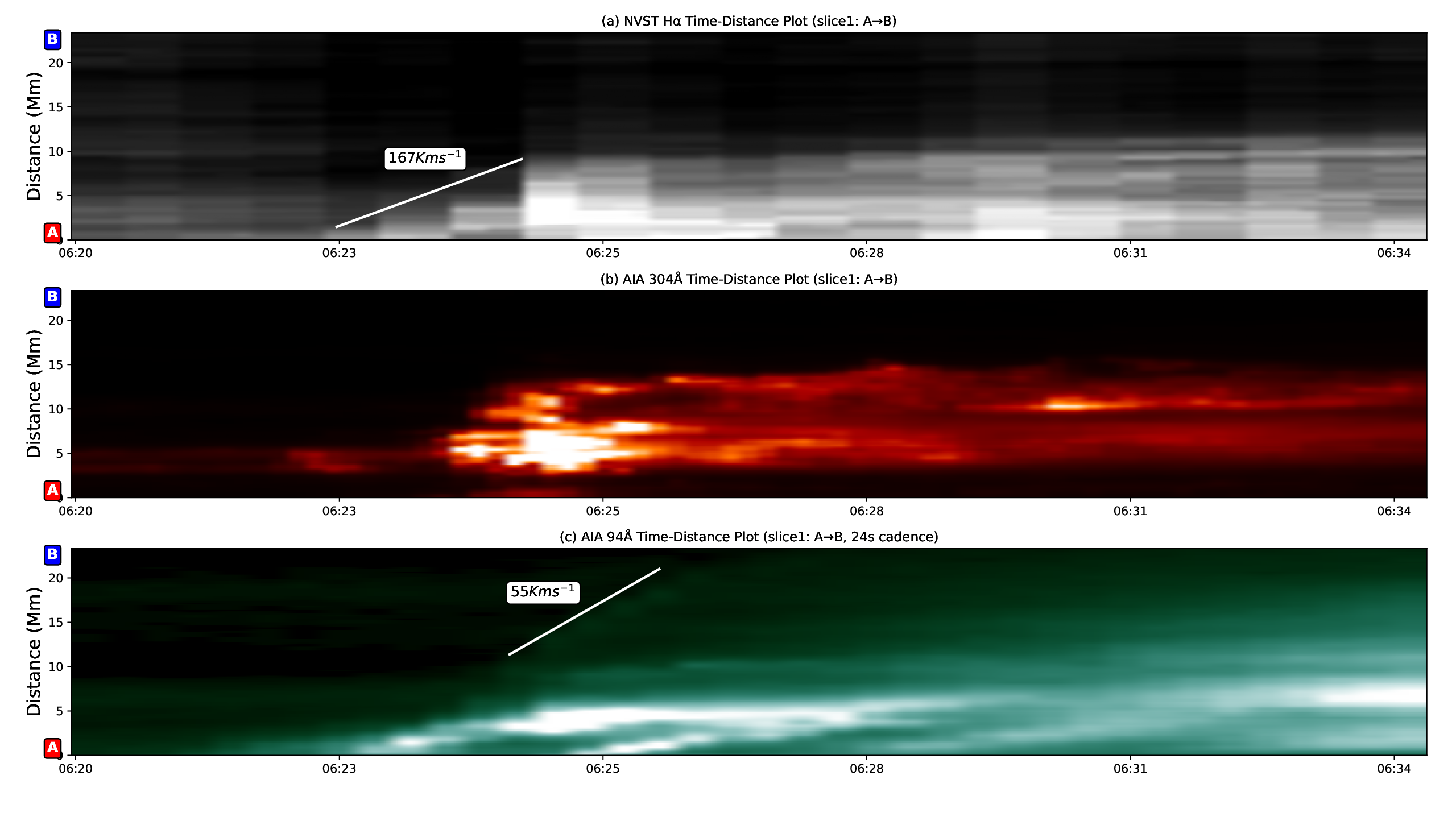}
\caption{Time--distance plots along the slits shown in Figure 1. (a) NVST H$\alpha$, (b) AIA 304~\AA, and (c) AIA 94~\AA\ along the same slit, shown as the blue line in Figure 1d and the white line AB in Figure 1g. Panels (a) and (b) present the extension of the chromospheric brightening region, while panel (c) shows the motion of the hot-channel structure. }
\label{fig:2}
\end{figure*}

\begin{figure*}
\plotone{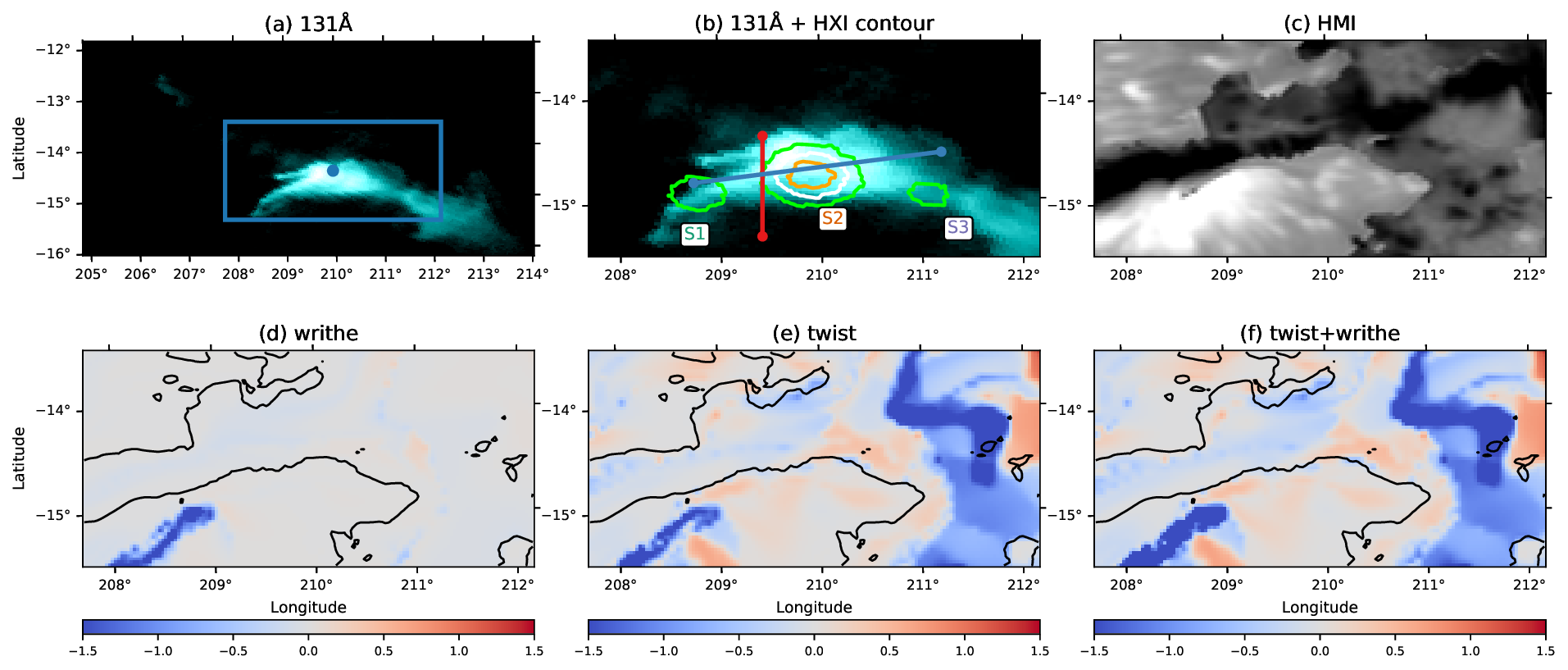}
\caption{ (a) AIA 131~\AA\ image showing the region of interest. (b) Enlarged view of the blue box in panel (a); overlaid are HXI contours in the 20--30~keV band, integrated from 06:28~UT to 06:30~UT and plotted from 98\% to 100\% of the peak intensity, with three contour levels separated by 1\%. (c) SDO/HMI radial magnetic field ($B_{r}$), used as the boundary condition for the NLFFF extrapolation. (d) Photospheric writhe distribution derived from Equation~(2). (e) Photospheric twist distribution derived from Equation~(3). (f) Photospheric helicity-proxy distribution derived from Equation~(1). 
In panels (d)--(f), the black contour indicates $B_{r}=0$, highlighting the relative spatial distributions of the different helicity-related quantities in the vicinity of the PIL.}
\label{fig:3}
\end{figure*}

\begin{figure*}
\plotone{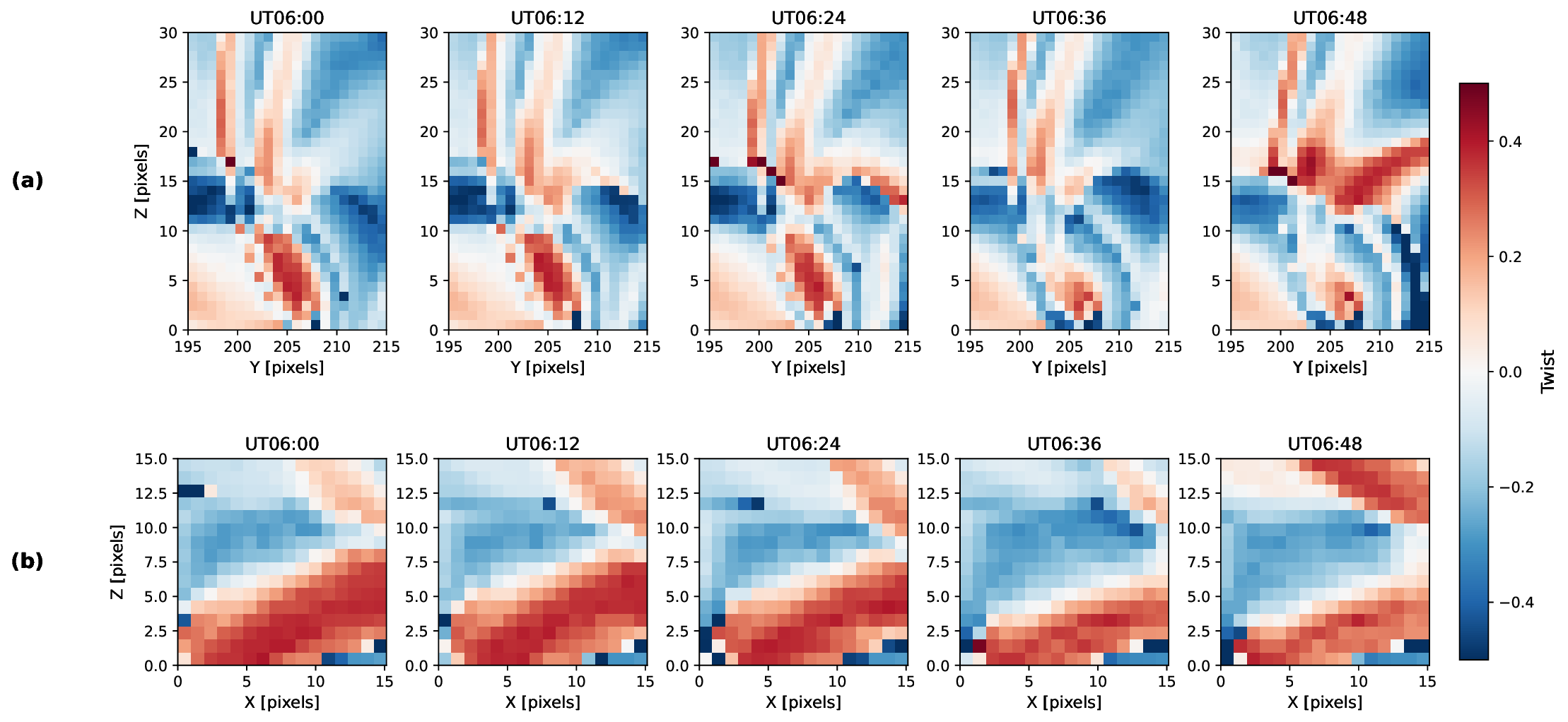}
\caption{Time evolution of the helicity-proxy distribution in two vertical cross-sections derived from the NLFFF extrapolation. The upper and lower rows correspond to the cross-sections along the red and blue lines marked in Figure 3b, respectively. From left to right, the columns show the distributions at 06:00, 06:12, 06:24, 06:36, and 06:48~UT. In each panel, the horizontal axis gives the distance along the corresponding line in Figure 3, and the vertical axis gives the height above the photosphere. Each pixel shows the value of the helicity proxy assigned to the magnetic field line traced from the corresponding point in the cross-section. The pixel size is 0.73 Mm.}
\label{fig:4}
\end{figure*}

\begin{figure*}
\plotone{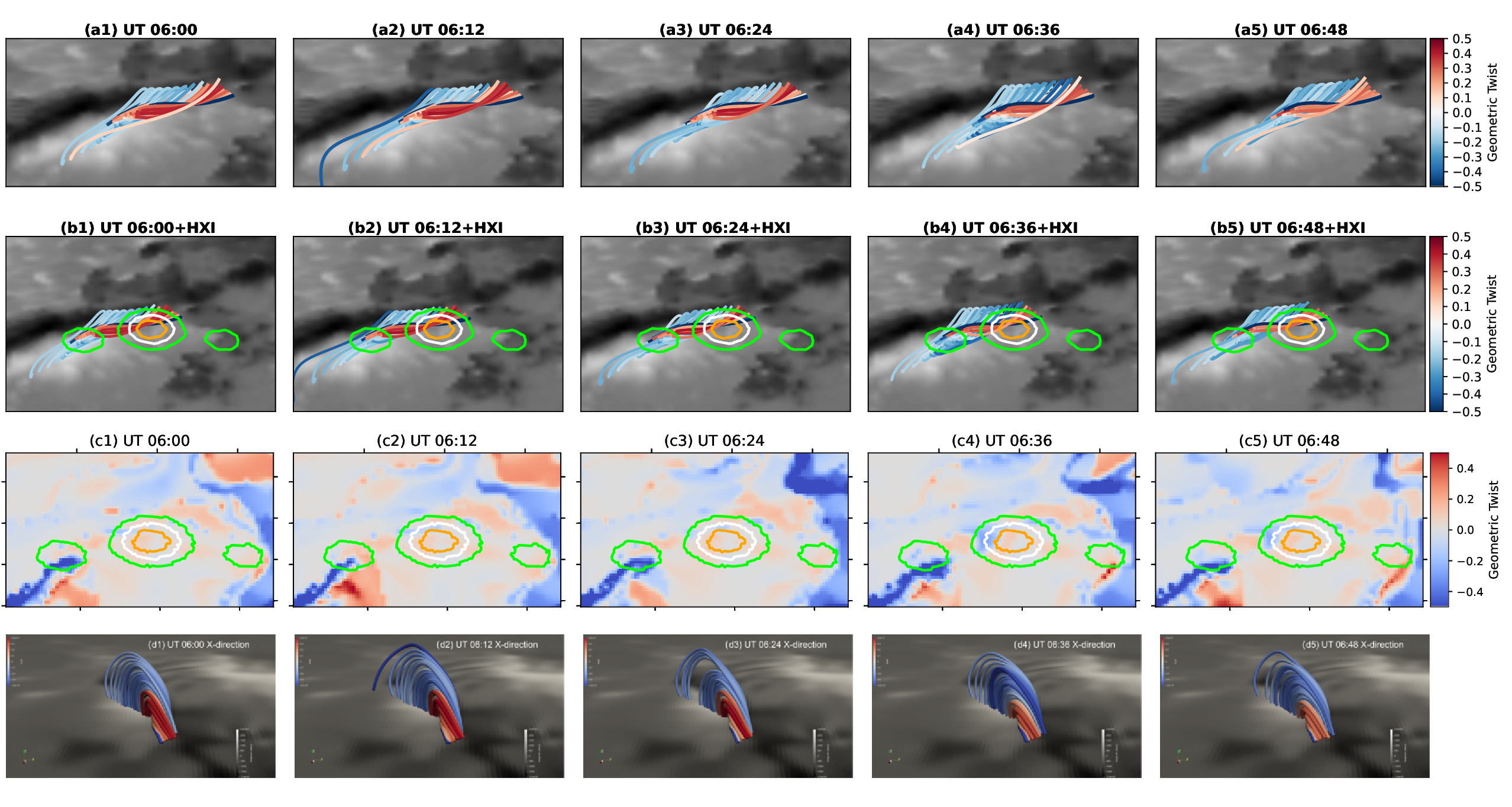}
\caption{Evolution of the extrapolated magnetic field lines and their relation to the HXI sources. Row (a) shows the magnetic field-line structure at five times from 06:00~UT to 06:48~UT. Row (b) shows the same field lines overlaid with HXI contours. Row (c) shows the corresponding bottom-boundary helicity distribution overlaid with HXI contours. Row (d) provides an enlarged view of the western footpoint region. The HXI contours are obtained in the 20--30~keV band, integrated from 06:28~UT to 06:30~UT, and plotted from 98\% to 100\% of the peak intensity, with three contour levels separated by 1\%.}
\label{fig:5}
\end{figure*}

\begin{figure*}
\plotone{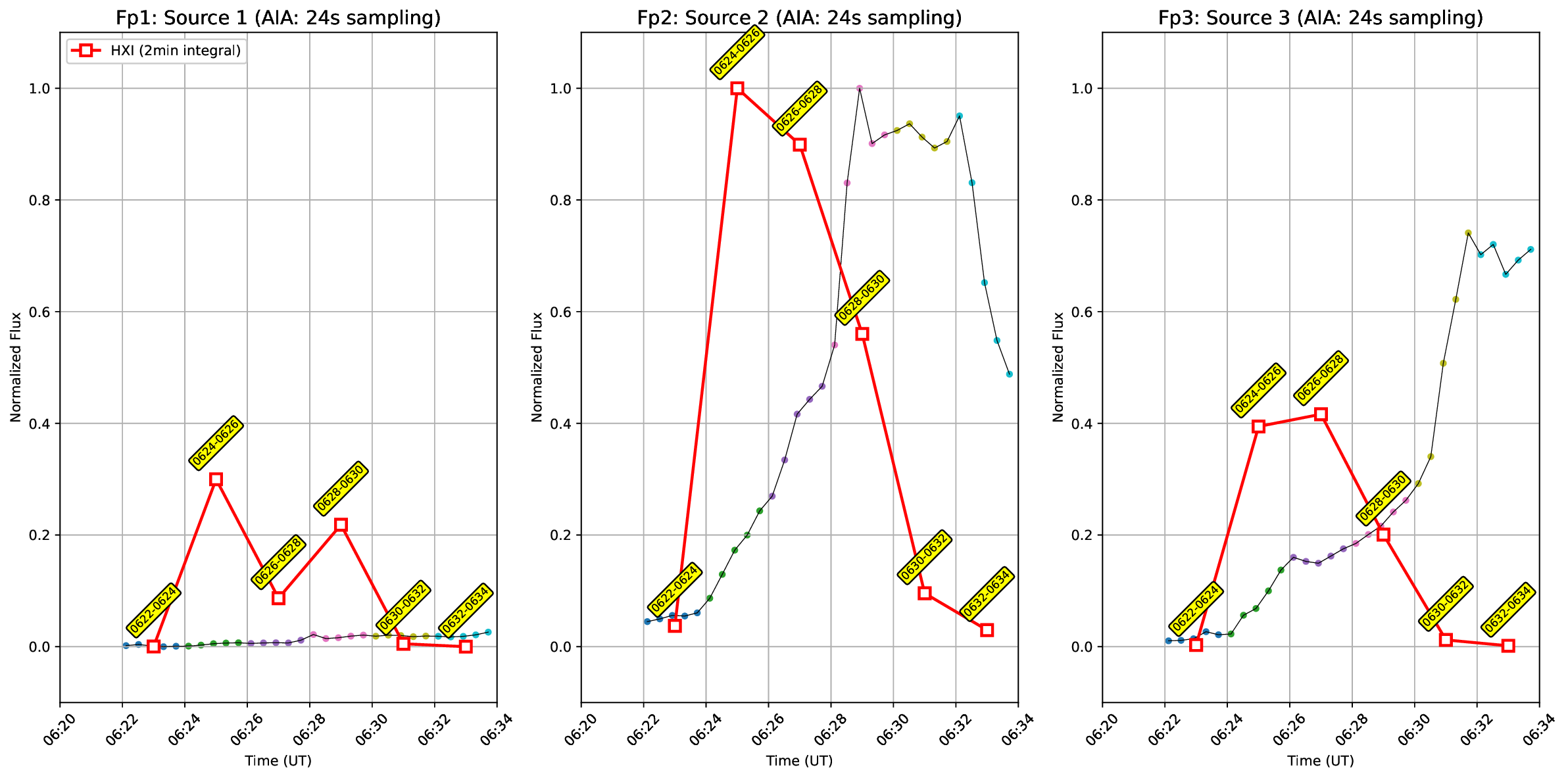}
\caption{Light curves of the three hard X-ray sources labeled in Figure 5(b) in the 20–30 keV band with a 1 minute cadence. The dotted lines indicate the intensity variations in the AIA 304~\AA\ channel from the corresponding regions.}
\label{fig:6}
\end{figure*}

\begin{figure*}
\plotone{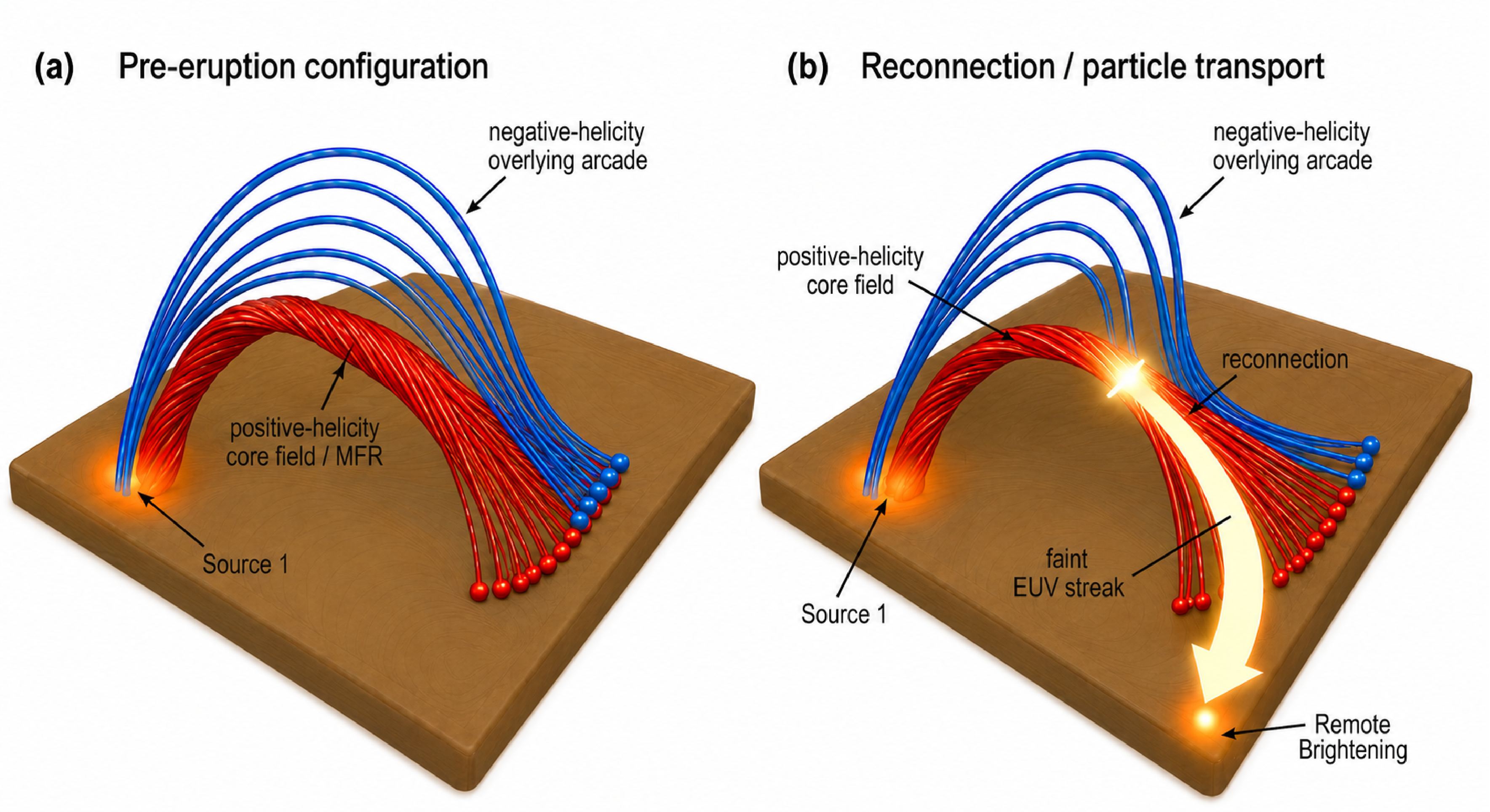}
\caption{Schematic illustration of the counter-helicity reconnection scenario. (a) Pre-eruption configuration showing a low-lying positive-helicity core field (red) interacting with an overlying negative-helicity arcade (blue). (b) Reconnection process and particle transport. The magnetic reconnection between the core field and the overlying arcade may contribute to the destabilization of the erupting core field, while accelerating particles or plasma that may contribute to the remote brightening and the EUV stripe-like structures.}
\label{fig:7}
\end{figure*}

\FloatBarrier

\begin{acknowledgments}
We are indebted to the SDO, NVST, ASO-S/HXI, CHASE, and GOES teams for providing the data. This work is supported by the Strategic Priority Research Program of the Chinese Academy of Sciences, grant No. XDB0560000, the National Natural Science Foundation of China under grants 12273106 and 12273108, the “Yunnan Revitalization Talent Support Program” Innovation Team Project (202405AS350012), the Yunnan Science Foundation of China (202501AT070025, 202501AT071627), and the Yunnan Province XingDian Talent Support Program. We thank the reviewers and editors for their valuable comments during the submission and revision process

\end{acknowledgments}





\bibliography{paper2final_clean}{}
\bibliographystyle{aasjournalv7}
\end{document}